\documentstyle[12pt]{article}
\def\beq{\begin{equation}}
\def\eeq{\end{equation}}

\begin{document}

\begin{titlepage}
\begin{flushright}
BA-00-06\\
\end{flushright}

\begin{center}
{\Large\bf   Anomalous Flavor ${\cal U}(1)$: Predictive Texture \\
For Bi-maximal Neutrino Mixing}
\end{center}
\vspace{0.5cm}
\begin{center}
{\large Qaisar Shafi$^{a}$\footnote {E-mail address:
shafi@bartol.udel.edu} {}~and
{}~Zurab Tavartkiladze$^{b}$\footnote {E-mail address:
z\_tavart@osgf.ge} }
\vspace{0.5cm}

$^a${\em Bartol Research Institute, University of Delaware,
Newark, DE 19716, USA \\

$^b$ Institute of Physics, Georgian Academy of Sciences,
380077 Tbilisi, Georgia}\\
\end{center}

\vspace{1.0cm}

\begin{abstract}

We present a scenario which naturally provides bi-maximal
neutrino mixings for a simultaneous explanation of the recent
atmospheric and solar neutrino data.
A crucial role is played by an anomalous flavor ${\cal U}(1)$
symmetry, which is also important for a natural understanding
of charged fermion mass hierarchies and magnitudes of
the CKM matrix
elements.
Within MSSM the solar neutrino problem can be resolved either
through vacuum oscillations or the large mixing angle MSW
solution. In supersymmetric GUTs such as $SU(5)$ and $SO(10)$
the MSW
solution is realized. If the flavor ${\cal U}(1)$ also
mediates supersymmetry breaking, the vacuum solution in MSSM
is eliminated, and only the large mixing angle MSW solution
survives.

\end{abstract}

\end{titlepage}


Within the various atmospheric and solar neutrino
oscillation solutions
allowed by the SuperKamiokande (SK) data (see \cite{atmSK}
and \cite{solSK} respectively), the
bi-maximal mixing scenario is particularly attractive and
much studied \cite{bimax}-\cite{moh}.
Recently an interesting and relatively simple texture
for the neutrino mass matrix

\begin{equation}
\begin{array}{ccc}
&  {\begin{array}{ccc}
\hspace{-5mm}~~ & \,\, & \,\,

\end{array}}\\ \vspace{2mm}
\begin{array}{c}
\\  \\
\end{array}\!\!\!\!\!\hat{M}_{\nu }=&{\left(\begin{array}{ccc}
\,\,0~~ &\,\,m_1~~ &
\,\,m_2
\\
\,\,m_1~~   &\,\,0~~  &
\,\,0
 \\
\,\,m_2~~ &\,\,0~~ &\,\,0
\end{array}\right) }~
\end{array}  \!\!~~~~~
\label{nu}
\end{equation}
has been proposed \cite{texture, moh}, which naturally
yields large
$\nu_{\mu }-\nu_{\tau }$ mixing (if mass scales $m_1, m_2$
are of the same order), and maximal $\nu_e-\nu_{\mu, \tau }$
oscillations. The $\nu_{\mu }-\nu_{\tau }$ mixing
can  become maximal
for $m_1\simeq m_2$. In \cite{texture} an analysis of
the texture in (\ref{nu})
was presented, but not how it was generated.
In \cite{moh} a specific model realizing
(\ref{nu}) within the standard model (MS) framework was given,
and requires the introduction of additional scalar
doublets as well as triplet. Suitable discrete and
continuous symmetries also must be imposed.
It seems natural to search for an alternative,
possibly a more economical framework for
implementing (\ref{nu}).

In this paper we present a simple way for realizing
(\ref{nu}) in which an
anomalous ${\cal U}(1)$ flavor symmetry
plays a crucial role. Atmospheric anomaly
is due to large $\nu_{\mu }-\nu_{\tau }$ oscillations.
Although, (\ref{nu}) gives maximal
$\nu_e-\nu_{\mu, \tau }$ mixing to explain the solar
neutrino deficit, a priori it is not clear
whether this corresponds to the vacuum or large
angle MSW oscillations.
As we show below, this depends on the specifics of the
scenario (since deviations from the zero entries in (\ref{nu}),
determining the splitting $\Delta m^2_{12}$, are model
dependent).
The ${\cal U}(1)$ flavor
symmetry also helps provide a natural understanding
of the hierarchies
between the charged fermion masses and
the CKM matrix elements.

Before presenting the mechanism, let us demonstrate how the
texture in (\ref{nu}) leads to bi-maximal mixings.
Using an orthogonal transformation

\beq
U_1^T\hat{M}_{\nu }U_1=\hat{M'}_{\nu }~,
\label{tran1}
\eeq
where
\beq
\begin{array}{ccc}
U_1=~~
\!\!\!\!\!\!\!\!&{\left(\begin{array}{ccc}
\,\,1 &\,\,~~0 &
\,\,~~0
\\
\,\,0  &\,\,~~~~c_{\theta }
&\,\,~~-s_{\theta }
\\
\,\, ~0 &\,\,~
s_{\theta }  &\,\,~~c_{\theta }
\end{array}\right)}
\end{array}~,
\label{u1}
\end{equation}
\beq
s_{\theta }\equiv \sin \theta~,~~~c_{\theta }\equiv \cos \theta~,~~~
\tan \theta =\frac{m_2}{m_1}~,
\label{angles}
\eeq
the neutrino mass matrix takes the degenerate form

\begin{equation}
\begin{array}{ccc}
&  {\begin{array}{ccc}
\hspace{-5mm}~~ & \,\, & \,\,

\end{array}}\\ \vspace{2mm}
\begin{array}{c}
 \\  \\
 \end{array}\!\!\!\!\!\hat{M'}_{\nu }= &{\left(\begin{array}{ccc}
\,\,0~~ &\,\,m~~ &
\,\,0
\\
\,\,m~~   &\,\,0~~  &
\,\,0
\\
\,\,0 &\,\,0~~ &\,\,0
\end{array}\right) }~,
\end{array}  \!\!  ~~~~~~~m=\sqrt{m_1^2+m_2^2}~.
\label{nu2}
\end{equation}
(\ref{nu2}) is diagonalized through a transformation with maximal
rotation angles

\beq
U_2^T\hat{M'}_{\nu }U_2\equiv \hat{M}_{\nu }^{\rm diag}=
{\rm Diag }(m~,~-m~,~0)~,
\label{tran3}
\eeq
where
\beq
\begin{array}{ccc}
U_2=~~
\!\!\!\!\!\!\!\!&{\left(\begin{array}{ccc}
\,\,\frac{1}{\sqrt{2}} &\,\,~~-\frac{1}{\sqrt{2}} &
\,\,~~0
\\
\,\,\frac{1}{\sqrt{2}} &\,\,~~~\frac{1}{\sqrt{2}}
&\,\,~~0
\\
\,\, ~0 &\,\,~0  &\,\,~~1
\end{array}\right) }
\end{array}~.
\label{u2}
\end{equation}
Taking into account (\ref{u1}), (\ref{u2}) the neutrino mixing matrix is

\beq
\begin{array}{ccc}
U_{\nu }=U_1U_2=~~
\!\!\!\!\!\!\!\!&{\left(\begin{array}{ccc}
\,\,\frac{1}{\sqrt{2}} &\,\,~~-\frac{1}{\sqrt{2}} &
\,\,~~0
\\
\,\,\frac{1}{\sqrt{2}}c_{\theta }  &\,\,~~~~\frac{1}{\sqrt{2}}c_{\theta }
&\,\,~~-s_{\theta }
\\
\,\, ~~\frac{1}{\sqrt{2}}s_{\theta } &\,\,~
\frac{1}{\sqrt{2}}s_{\theta }  &\,\,~~c_{\theta }
\end{array}\right) }
\end{array}~.
\label{lepckm}
\end{equation}
{}From (\ref{lepckm}) [taking into account (\ref{angles})] the
atmospheric and solar neutrino oscillation amplitudes respectively
are

$$
{\cal A}(\nu_{\mu }\to \nu_{\tau })=
\frac{4m_1^2m_2^2}{(m_1^2+m_2^2)^2}~~,
$$
\beq
{\cal A}(\nu_e\to \nu_{\mu, \tau })=1~,
\label{amp}
\eeq
where the oscillation amplitudes are defined as

\beq
{\cal A}(\nu_{\alpha }\to \nu_{\beta })=
4\Sigma_{j<i}U_{\nu }^{\alpha j}U_{\nu }^{\alpha i}
U_{\nu }^{\beta j}U_{\nu }^{\beta i}~,
\label{defamp}
\eeq
($\alpha, \beta $ denote flavor indices and $i, j$
the mass eigenstates).

{}From (\ref{amp}) we see that for $m_1\simeq m_2$, we have
the bi-maximal oscillations scenario. As far as the neutrino
${\rm mass}^2$ splittings are concerned, since the mass spectrum
is

\beq
m_{\nu_1}=m_{\nu_2}=m~,~~~~~~~~m_{\nu_3}=0~,
\label{masses}
\eeq
we will have

\beq
\Delta m_{32}^2=m^2~,~~~~\Delta m_{21}^2=0~.
\label{split}
\eeq
Having $m^2\sim m^2_{\rm atm}\sim 10^{-3}~{\rm eV}^2$, the atmospheric
neutrino puzzle is successfully resolved.
However, without a
non-zero $\Delta m_{21}^2=0$, the solar neutrino oscillations
will be absent. This deviation from zero will determine whether
the vacuum or MSW solution is realized. In
\cite{moh} the non-zero splittings emerge from radiative
corrections, and led to the vacuum oscillation solution.
We will discuss this issue within the framework of
MSSM as well as SUSY $SU(5)$.
We also indicate the implications if the anomalous ${\cal U}(1)$
mediates in addition SUSY breaking.

\section{Model with ${\cal U}(1)$ flavor symmetry}

We introduce an anomalous ${\cal U}(1)$ flavor symmetry which,
may arise
in effective field theories from strings.
The cancellation of its anomalies occurs through the Green-Schwarz
mechanism
\cite{gsh}. Due to the anomaly, the Fayet-Iliopoulos term

\beq
\xi \int d^4\theta V_A
\label{fi}
\eeq
is always generated \cite{fi}, where, in string theory, $\xi $ is given by
\cite{xi}

\begin{equation}
\xi =\frac{g_A^2M_P^2}{192\pi^2}{\rm Tr}Q~.
\label{xi}
\end{equation}
The $D_A$-term will have the form
\begin{equation}
\frac{g_A^2}{8}D_A^2=\frac{g_A^2}{8}
\left(\Sigma Q_a|\varphi_a |^2+\xi \right)^2~,
\label{da}
\end{equation}
where $Q_a$ is the `anomalous' charge of $\varphi_a $ superfield.

In \cite{gia} the anomalous ${\cal U}(1)$ symmetry was considered as
a mediator of SUSY breaking. In \cite{anu1}, the anomalous
Abelian
symmetries were exploited as flavor symmetries for a natural understanding
of hierarchies of fermion masses and mixings,
while in \cite{nuu1} the various neutrino oscillation scenarios with
${\cal U}(1)$
symmetry were studied.

Assuming ${\rm Tr}Q>0$ ($\xi >0$) and introducing a superfield
$X$ with $Q_{X}=-1$,
we can ensure that the cancellation of (\ref{da})
fixes the VEV of the scalar component of $X$ to be

\beq
\langle X\rangle =\sqrt{\xi }~.
\label{vevx}
\eeq
Further, we will assume that

\beq
\frac{\langle X\rangle }{M_P}\equiv \epsilon \simeq 0.2~.
\label{epsx}
\eeq
The parameter $\epsilon $ is an important expansion parameter for
understanding the magnitudes of
fermion masses and mixings.

Starting our considerations with the neutrino sector
within the framework of MSSM (which is more
general than some specific GUT model), let us consider the following
prescription for the ${\cal U}(1)$ charges

\beq
Q_X=-1~,~Q_{l_2}=Q_{l_3}=k~,~Q_{l_1}=k+n~,~Q_{h_u}=Q_{h_d}=0~.
\label{charges}
\eeq
In order to obtain non-zero neutrino masses, we introduce
two right-handed neutrino superfields ${\cal N}_1, {\cal N}_2$,
which, by a judicious choice of ${\cal U}(1)$ charges
\footnote{For models in which ${\cal U}(1)$ flavor symmetry
plays a crucial role for achieving maximal/large mixings,
see refs. \cite{maxmix, ourbimax}.}

\beq
Q_{{\cal N}_1}=-Q_{{\cal N}_2}=k~,
\label{nucharges}
\eeq
will provide a texture similar to (\ref{nu}).
{}From (\ref{charges}), (\ref{nucharges}), the
relevant couplings will be:

\begin{equation}
\begin{array}{cc}
 & {\begin{array}{cc}
{\cal N}_1~&\,\,{\cal N}_2~~~~~~
\end{array}}\\ \vspace{2mm}
\begin{array}{c}
l_1\\ l_2 \\ l_3

\end{array}\!\!\!\!\! &{\left(\begin{array}{ccc}
\,\, \epsilon^{2k+n}~~ &
\,\, \epsilon^{n}
\\
\,\, \epsilon^{2k} ~~ &\,\,0
\\
\,\, \epsilon^{2k} ~~ &\,\,0
\end{array}\right)h_u }~,
\end{array}  \!\!~~~
\begin{array}{cc}
 & {\begin{array}{cc}
{\cal N}_1~&\,\,
{\cal N}_2~~~~~
\end{array}}\\ \vspace{2mm}
\begin{array}{c}
{\cal N}_1 \\ {\cal N}_2

\end{array}\!\!\!\!\! &{\left(\begin{array}{ccc}
\,\, \epsilon^{2k}
 &\,\,~~~1
\\
\,\, 1
&\,\,~~~0
\end{array}\right)M~,
}
\end{array}~~~
\label{Ns}
\end{equation}
where $M$ is some mass scale. Integration of ${\cal N}_{1, 2}$
states yields the neutrino mass matrix

\begin{equation}
\begin{array}{ccc}
&  {\begin{array}{ccc}
\hspace{-5mm}~~ & \,\, & \,\,

\end{array}}\\ \vspace{2mm}
\begin{array}{c}
 \\  \\
 \end{array}\!\!\!\!\!\hat{M}_{\nu }\propto &{\left(\begin{array}{ccc}
\,\,\epsilon^n~~ &\,\,1~~ &
\,\,1
\\
\,\,1~~   &\,\,0~~  &
\,\,0
 \\
\,\,1~~ &\,\,0~~ &\,\,0
\end{array}\right) }m~,~~~~~~m=\frac{\epsilon^{2k+n}h_u^2}{M}~,
\end{array}  \!\!
\label{u1nu}
\end{equation}
which resembles the texture (\ref{nu}), but differs from it by a
nonzero (1,1) entry, and provides the $\Delta m_{21}^2$ splitting.

{}From (\ref{u1nu}) and (\ref{defamp}) we have for
the atmospheric and solar
neutrino
oscillation parameters (respectively)

$$
\Delta m^2_{32}\equiv m_{\rm atm}^2= m^2\sim 10^{-3}~{\rm eV}^2~,
$$
\beq
{\cal A}(\nu_{\mu }\to \nu_{\tau })\sim 1~,
\label{atmosc}
\eeq

$$
\Delta m^2_{21 }\simeq 2m_{\rm atm}^2\epsilon^n~,
$$
\beq
{\cal A}(\nu_e \to \nu_{\mu , \tau }) =1-{\cal O}(\epsilon^{2n})~.
\label{solosc}
\eeq
We observe that the ${\rm mass}^2$ splitting for solar neutrinos is
expressed in terms of the atmospheric scale $m_{\rm atm}$
and $n$-th power of
$\epsilon $.

{}From (\ref{solosc}) we have

\beq
\Delta m_{21}^2\propto \left\{ \begin{array}{ll}
10^{-10}~{\rm eV}^2 & \mbox{for $n=10$} \\
10^{-5}~{\rm eV}^2 & \mbox{for $n=3$}
\end{array}
\right.
\label{masssol}
\eeq
Therefore, $n=10$ corresponds to vacuum oscillations
of solar neutrinos, while $n=3$ gives the large angle MSW solution.
The MSSM does not constrain $n$ to be either $10$ or $3$, and so
both scenarios are possible. To see this, let us consider
the charged fermion sector. With the prescription

\beq
Q_{e_3^c}=p~,~~~Q_{e_2^c}=p+2~,~~~
Q_{e_1^c}=p+5-n~,~~~
\label{erch}
\eeq
the Yukawa couplings for charged leptons have the form

\begin{equation}
\begin{array}{ccc}
 & {\begin{array}{ccc}
\hspace{-5mm} e^c_1~ & \,\,e^c_2 ~~ & \,\,e^c_3 ~~

\end{array}}\\ \vspace{2mm}
\begin{array}{c}
l_1 \\ l_2 \\l_3
 \end{array}\!\!\!\!\! &{\left(\begin{array}{ccc}
\,\,\epsilon^5~~ &\,\,\epsilon^{n+2}~~ &
\,\,\epsilon^n
\\
\,\,\epsilon^{5-n}~~   &\,\,\epsilon^2~~  &
\,\,1
 \\
\,\,\epsilon^{5-n}~~ &\,\,\epsilon^2~~ &\,\,1
\end{array}\right)\epsilon^{p+k}h_d }~,
\end{array}  \!\!  ~~~~~
\label{down}
\end{equation}
providing the desirable hierarchies

\beq
\lambda_{\tau }\sim \epsilon^{p+k}~,~~
\lambda_e :\lambda_{\mu } :\lambda_{\tau } \sim
\epsilon^5:\epsilon^2 :1~,
\label{lambdas}
\eeq
with

\beq
\tan \beta \sim \epsilon^{p+k}\frac{m_t}{m_b}~.~~
\label{tanbeta}
\eeq

As far as the quark sector is concerned, with

$$
Q_{q_3}=0~,~~Q_{q_2}=2~,~~Q_{q_1}=3~,~~
Q_{d^c_3}=Q_{d^c_2}=p+k~,~~
$$
\beq
Q_{d^c_1}=p+k+2~,~~Q_{u^c_3}=0~,~~Q_{u^c_2}=1~,~~Q_{u^c_1}=3~,
\label{qch}
\eeq
the appropriate Yukawa couplings are

\begin{equation}
\begin{array}{ccc}
 & {\begin{array}{ccc}
\hspace{-5mm} d^c_1~ & \,\,d^c_2 ~~ & \,\,d^c_3 ~~

\end{array}}\\ \vspace{2mm}
\begin{array}{c}
q_1 \\ q_2 \\q_3
 \end{array}\!\!\!\!\! &{\left(\begin{array}{ccc}
\,\,\epsilon^5~~ &\,\,\epsilon^3~~ &
\,\,\epsilon^3
\\
\,\,\epsilon^4~~   &\,\,\epsilon^2~~  &
\,\,\epsilon^2
 \\
\,\,\epsilon^2~~ &\,\,1~~ &\,\,1
\end{array}\right)\epsilon^{p+k}h_d }~,
\end{array}  \!\!  ~~~~~
\label{down}
\end{equation}

\begin{equation}
\begin{array}{ccc}
 & {\begin{array}{ccc}
\hspace{-5mm} u^c_1~ & \,\,u^c_2 ~~ & \,\,u^c_3 ~~

\end{array}}\\ \vspace{2mm}
\begin{array}{c}
q_1 \\ q_2 \\q_3
 \end{array}\!\!\!\!\! &{\left(\begin{array}{ccc}
\,\,\epsilon^6~~ &\,\,\epsilon^4~~ &
\,\,\epsilon^3
\\
\,\,\epsilon^5~~   &\,\,\epsilon^3~~  &
\,\,2
 \\
\,\,\epsilon^3~~ &\,\,\epsilon ~~ &\,\,1
\end{array}\right)h_u }~,
\end{array}  \!\!  ~~~~~
\label{up}
\end{equation}
yielding

\beq
\lambda_b\sim \epsilon^{p+k}~,~~
\lambda_d :\lambda_s :\lambda_b \sim
\epsilon^5:\epsilon^2 :1~,
\label{dlambdas}
\eeq

\beq
\lambda_t\sim 1~,~~
\lambda_u :\lambda_c :\lambda_t \sim
\epsilon^6:\epsilon^3 :1~.
\label{ulambdas}
\eeq

{}From (\ref{down}), (\ref{up}), for the CKM matrix elements
we find

\beq
V_{us}\sim \epsilon~,~~~V_{cb}\sim \epsilon^2~,~~~
V_{ub}\sim \epsilon^3~.
\label{ckm}
\eeq

We see that the MSSM does not fix the values of $n, p, k$. 
However, specific GUTs can be
more restrictive. To demonstrate this,
we consider the simplest version of $SU(5)$ GUT,
with three families of $(10+\bar 5)$-plets.
Due to these unified multiplets:


\beq
Q_q=Q_{e^c}=Q_{u^c}=Q_{10}~,~~~~~
Q_l=Q_{d^c}=Q_{\bar 5}~.
\label{chsu5}
\eeq
The known hierarchies (\ref{ckm}) of the CKM matrix elements
now fix the relative charges of $10$-plets,

\beq
Q_{10_3}=0~,~~~Q_{10_2}=2~,~~~Q_{10_1}=3~,
\label{ch10}
\eeq
while (\ref{lambdas}), (\ref{dlambdas}) dictate

\beq
Q_{\bar 5_3}=Q_{\bar 5_2}=k~,~~~~Q_{\bar 5_1}=k+2~.
\label{ch5}
\eeq
Comparing (\ref{chsu5})-(\ref{ch5}) with (\ref{charges}), (\ref{erch})
(\ref{qch}) we see that the minimal $SU(5)$ GUT fixes $n$ and $p$
to be

\beq
n=2~,~~~~~p=0~,
\label{nk}
\eeq
{}From (\ref{solosc}) (which now turns out to be predictive since
$n$ is fixed) we get

\beq
\Delta m_{21}^2\sim 10^{-4}~{\rm eV}^2~.
\label{solsu5}
\eeq
This value is close to the scale corresponding to the large
angle MSW oscillations of the solar neutrinos. We see that our
mechanism within $SU(5)$ GUT strongly suggests large angle MSW
oscillations for solar neutrinos.

The same conclusion can be reached with $SO(10)$
GUT, since also in this case the prescription of the ${\cal U}(1)$
charges \cite{ourbimax} would exclude the vacuum solution of the
solar neutrino problem.
Within this framework the
large $\nu_{\mu }-\nu_{\tau }$ mixing remains unchanged.

Let us note that the conclusions presented above are valid if
the anomalous ${\cal U}(1)$ only acts as flavor symmetry
and is not tied
with SUSY breaking. In several models an anomalous ${\cal U}(1)$
symmetry also acts as a mediator of SUSY breaking \cite{gia}.
This can be very useful for adequate suppression
of FCNC \cite{dec, nucl} and dimension five nucleon decay \cite{nucl}.
In this type of scenarios the soft ${\rm mass}^2$ for sparticles, which
have non-zero ${\cal U}(1)$ charges, emerge from non-zero
$D_A$-term and equal

\beq
m^2_{\tilde{\phi}_i}=m_S^2Q_{\phi_i}~,
\label{soft}
\eeq
where $m_S$ is taken ${\cal O}(10~{\rm TeV})$. Therefore,
the ${\cal U}(1)$
charges of matter superfields must be positive in order to avoid
$SU(3)_c\times U(1)_{\rm em}$ breaking. On the other hand, from
(\ref{erch}) we see that the choice $n=10$ is excluded (!)
[since $0\stackrel{<}{_{-}}p+k\stackrel{<}{_{-}}3$
($1\stackrel{>}{_\sim }\lambda_{\tau, b }\stackrel{<}{_\sim } 10^{-2}$)],
which means that in this case the vacuum oscillation solution
for solar neutrinos is not realized even within the MSSM framework.
The cases $n=2, 3$, which correspond to the large angle MSW
solution, are still allowed.
The interesting point is that the neutrino oscillation
scenario is linked with SUSY breaking mechanism.

In conclusion, we have suggested a scenario
for obtaining bi-maximal neutrino mixing. A crucial role is
played by an anomalous ${\cal U}(1)$ flavor symmetry
for obtaining the simple neutrino mass matrix texture in (\ref{nu}).
The atmospheric neutrino puzzle is resolved by large/maximal
$\nu_{\mu}-\nu_{\tau }$ mixing, while the scenario for large angle
$\nu_e-\nu_{\mu, \tau }$ oscillations is model dependent:
within the MSSM, both the large angle vacuum or the large angle MSW
oscillations are possible, while the $SU(5)$ GUT
(and also $SO(10)$)
predicts the
large angle MSW solution.
If the anomalous ${\cal U}(1)$ flavor symmetry is
also a mediator of SUSY breaking, then the solar neutrino
vacuum oscillations are excluded
and the large angle MSW solution is responsible for
the solar neutrino deficit.
Finally, the ${\cal U}(1)$ flavor symmetry also nicely explains
the hierarchies between the charged fermion masses and
the magnitudes of the CKM matrix elements.

\vspace{0.2cm}

This work was supported in part by the DOE under Grant No.
DE-FG02-91ER40626.

\end{document}